\documentclass[nonacm,sigconf,fleqn,svgnames,screen]{acmart}

\AtBeginDocument{%
	\providecommand\BibTeX{{%
			\normalfont B\kern-0.5em{\scshape i\kern-0.25em b}\kern-0.8em\TeX}}}

\setcopyright{none}
\copyrightyear{none}
\acmYear{none}
\acmDOI{none}

\definecolor{ml2rblu}{rgb}{0.02,0.27,0.45}
\definecolor{ml2ryel}{rgb}{0.98,0.72,0.18}
\definecolor{ml2rgrn}{rgb}{0.50,0.71,0.18}
\definecolor{ml2rtrq}{rgb}{0.00,0.57,0.57}

\usepackage{amsthm}
\usepackage{mathtools}
\usepackage{bm}
\usepackage[font=footnotesize]{subfig}
\usepackage{booktabs}
\usepackage{array}
\usepackage[inline]{enumitem}
\usepackage{fixltx2e}
\usepackage{nicefrac}
\usepackage[capitalize]{cleveref}

\usepackage{wasysym}
\usepackage{xspace}

\usepackage{tikz}
\usetikzlibrary{shapes.geometric, positioning, calc}

\usepackage{algorithm}
\usepackage[noend]{algpseudocode}

\usepackage{hyperref}
\hypersetup{%
	colorlinks=true,
	urlcolor=ml2rtrq,
	linkcolor=ml2rtrq,
	citecolor=ml2rtrq,
	bookmarks=false}

\usepackage{fancyvrb} 
\usepackage{listings}
\lstdefinestyle{pythonstyle}{%
	language=Python,
	tabsize=4,
	backgroundcolor=\color{Gray!10},
	basicstyle=\ttfamily\scriptsize,
	stringstyle=\color{ForestGreen},
	keywordstyle=\color{BlueViolet},
	commentstyle=\itshape\color{DarkRed!90},
	identifierstyle=,
	emphstyle=\color{Blue},
	frame=lines,	
	showstringspaces=false,
	morekeywords={range, len, self, other, lambda, from, import, as, False, True, 
		enumerate, xrange, map, list, set, float, int, min, max, sorted, None},
	fancyvrb=true,
}
\lstnewenvironment{python}[1][]{\lstset{style=pythonstyle,#1}}{}

\lstdefinestyle{cstyle}{%
	language=C,
	tabsize=4,
	backgroundcolor=\color{Gray!10},
	basicstyle=\ttfamily\scriptsize,
	stringstyle=\color{ForestGreen},
	keywordstyle=\color{BlueViolet},
	commentstyle=\itshape\color{DarkRed!90},
	identifierstyle=,
	emphstyle=\color{Blue},
	frame=lines,	
	showstringspaces=false,
	morekeywords={},
	fancyvrb=true,
}
\lstnewenvironment{ccode}[1][]{\lstset{style=cstyle,#1}}{}

\lstdefinestyle{pythonstyletxt}{%
	language=Python,
	tabsize=4,
	basicstyle=\ttfamily\small,
	stringstyle=\color{ForestGreen},
	keywordstyle=\color{BlueViolet},
	commentstyle=\itshape\color{DarkRed!90},
	identifierstyle=,
	emphstyle=\color{Blue},
	xleftmargin=1em,
	showstringspaces=false,
	morekeywords={range, len, self, other, lambda, from, import, as, False, True, 
		enumerate, xrange, map, list, set, float, int, min, max, sorted, with, None},
}
\lstnewenvironment{pythontxt}[1][]{\lstset{style=pythonstyletxt,#1}}{}

\lstdefinestyle{pythonstyletxtsmall}{%
	language=Python,
	tabsize=4,
	basicstyle=\ttfamily\scriptsize,
	stringstyle=\color{ForestGreen},
	keywordstyle=\color{BlueViolet},
	commentstyle=\itshape\color{DarkRed!90},
	identifierstyle=,
	emphstyle=\color{Blue},
	xleftmargin=1em,
	showstringspaces=false,
	morekeywords={range, len, self, other, lambda, from, import, as, False, True, 
		enumerate, xrange, map, list, set, float, int, min, max, sorted, None},
}
\lstnewenvironment{pythontxtsmall}[1][]{\lstset{style=pythonstyletxtsmall,#1}}{}

\newcommand{\putORCID}[1]{
	\authornote{\href{https://orcid.org/#1}{\includegraphics[width=2ex]{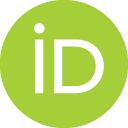}} \href{https://orcid.org/#1}{#1}}
	\orcid{#1}}

\usepackage{tikz}
\usetikzlibrary{shapes.geometric, positioning}
\usepackage{datatool}

\renewcommand{\vec}[1]{\bm{#1}}
\newcommand{\mat}[1]{\bm{#1}}
\newcommand{\diag}[1]{\operatorname{diag} \bigl [ #1 \bigr ]}

\newcommand{\BB}{\mathbb{B}}
\newcommand{\QUBO}{\textsf{QUBO}\xspace}
\newcommand{\tp}{\intercal}

\newcommand*{\defeq}{\mathrel{\vcenter{\baselineskip0.5ex\lineskiplimit0pt\hbox{\scriptsize.}\hbox{\scriptsize.}}}%
	=}
\newcommand*{\eqdef}{=\mathrel{\vcenter{\baselineskip0.5ex\lineskiplimit0pt\hbox{\scriptsize.}\hbox{\scriptsize.}}}}

\DeclarePairedDelimiter\abs\lvert\rvert
\DeclarePairedDelimiter\floor\lfloor\rfloor

\DTLloaddb[noheader=false]{nyts}{nyt-20240108-hard.csv}
\DTLloaddb[noheader=false]{nyts-solved}{nyt-20240108-hard-solution.csv}

\newcommand{\sudoku}[2]{%
	\begin{tikzpicture}[scale=#2,
		grid0/.style={draw=lightgray},
		grid1/.style={draw=lightgray,ultra thick},
		border/.style={ultra thick}]
		\foreach \l in {1,2,4,5,7,8}{
			\draw[grid0] ($(-0.5,\l-0.5)$) -- ++(9,0);
			\draw[grid0] ($(\l-0.5,-0.5)$) -- ++(0,9);}
		\foreach \l in {3,6}{
			\draw[grid1] ($(-0.5,\l-0.5)$) -- ++(9,0);
			\draw[grid1] ($(\l-0.5,-0.5)$) -- ++(0,9);}
		\draw[border] (-0.5,-0.5) -- ++(9,0) -- ++(0,9) -- ++(-9,0) -- ++(0,-9);
		\DTLforeach*{#1}{\i=i, \j=j, \k=k}{%
			\ifthenelse{\k<0}{%
				\def\col{ml2rgrn!75!white}
				\pgfmathsetmacro{\newk}{int(-\k)}
			}{%
				\def\col{black}
				\pgfmathsetmacro{\newk}{int(\k)}}
			\node[color=\col] at ($(\j, 8-\i)$) {\Huge \newk};}
\end{tikzpicture}}

\begin{document}
\begin{filecontents*}{nyt-20240108-hard.csv}
i,j,k
0,7,2
1,4,9
1,7,4
2,3,3
2,5,2
2,8,7
3,0,6
3,2,5
3,3,1
3,8,8
4,3,7
4,5,5
4,6,1
5,1,3
5,3,2
5,8,5
6,1,9
6,4,7
6,8,3
7,1,8
7,7,9
8,2,4
8,5,6
8,6,5
\end{filecontents*}

\begin{filecontents*}{nyt-20240108-hard-solution.csv}
i,j,k
0,7,2
1,4,9
1,7,4
2,3,3
2,5,2
2,8,7
3,0,6
3,2,5
3,3,1
3,8,8
4,3,7
4,5,5
4,6,1
5,1,3
5,3,2
5,8,5
6,1,9
6,4,7
6,8,3
7,1,8
7,7,9
8,2,4
8,5,6
8,6,5
0,0,-7
0,1,-1
0,2,-3
0,3,-8
0,4,-5
0,5,-4
0,6,-6
0,8,-9
1,0,-8
1,1,-5
1,2,-2
1,3,-6
1,5,-7
1,6,-3
1,8,-1
2,0,-4
2,1,-6
2,2,-9
2,4,-1
2,6,-8
2,7,-5
3,1,-4
3,4,-3
3,5,-9
3,6,-2
3,7,-7
4,0,-9
4,1,-2
4,2,-8
4,4,-6
4,7,-3
4,8,-4
5,0,-1
5,2,-7
5,4,-4
5,5,-8
5,6,-9
5,7,-6
6,0,-2
6,2,-6
6,3,-5
6,5,-1
6,6,-4
6,7,-8
7,0,-5
7,2,-1
7,3,-4
7,4,-2
7,5,-3
7,6,-7
7,8,-6
8,0,-3
8,1,-7
8,3,-9
8,4,-8
8,7,-1
8,8,-2
\end{filecontents*}

\hypersetup{%
  pdftitle={A Simple QUBO Formulation of Sudoku},
  pdfauthor={Sascha M\"ucke},
  pdfsubject={optimization, qubo, sudoku},
  pdfkeywords={optimization, qubo, sudoku}
}

\title[Sudoku QUBO]{A Simple QUBO Formulation of Sudoku}
\author[S. M\"ucke]{Sascha M\"ucke}
\putORCID{0000-0001-8332-6169}
\affiliation{
  \institution{Lamarr Institute\\ TU Dortmund University}
  \city{Dortmund}
  \state{Germany}
}

\begin{abstract}
This article describes how to solve Sudoku puzzles using Quadratic Unconstrained Binary Optimization (QUBO).
To this end, a QUBO instance with 729 variables is constructed, encoding a Sudoku grid with all constraints in place, which is then partially assigned to account for clues.
The resulting instance can be solved with a Quantum Annealer, or any other strategy, to obtain the fully filled-out Sudoku grid.
Moreover, as all valid solutions have the same energy, the QUBO instance can be used to sample uniformly from the space of valid Sudoku grids.
We demonstrate the described method using both a heuristic solver and a Quantum Annealer.
\end{abstract}

\maketitle

\section{Introduction}
\label{sec:intro}

Sudoku is a puzzle game originating from Japan, consisting of a $9\times 9$ square grid of cells that can hold the numbers 1 to 9.
Initially, some cells contain ``clues'', but most cells are empty.
The objective is to fill in the missing numbers in such a way that \begin{enumerate*}[label=(\roman*)]
\item each row contains all numbers exactly once,
\item each column contains all numbers exactly once,
\item each $3\times 3$ sub-grid (``block'') contains all numbers exactly once.
\end{enumerate*}
It has been shown that the problem is NP-complete for general grids of size $N^2\times N^2$ \cite{yato2003}.

\emph{Quadratic Unconstrained Binary Optimization} (\QUBO) is the problem of finding a binary vector $\vec x^*\in\BB^n$ with $\BB\defeq\lbrace 0,1\rbrace$ that minimizes the \emph{energy} function \begin{equation}\label{eq:qubo}
    f_{\mat Q}(\vec x)\defeq \vec x^{\tp}\mat Q\vec x=\sum_{\substack{i,j\in [n]\\ i\leq j}}Q_{ij}x_ix_j\;,
\end{equation}%
where $\mat Q\in\mathbb{R}^{n\times n}$ is an upper triangular \emph{weight} matrix, and $[n]$ denotes the set $\lbrace 1,\dots,n\rbrace$.
It is an \textsf{NP}-hard optimization problem \cite{pardalos1992} with numerous applications, ranging from economics \cite{laughhunn1970,hammer1971} over satisfiability \cite{kochenberger2005} and resource allocation \cite{neukart2017,chai2023} to Machine Learning \cite{bauckhage2018,mucke2019,bauckhage2020,date2020,bauckhage2021,piatkowski2022}, among others.
In recent years it has gained renewed attention because it is equivalent to the Ising model, which can be solved physically through quantum annealing \cite{kadowaki1998,farhi2000}, for which specialized quantum computers have been developed \cite{d-wavesystems2021}.
Aside from quantum annealing, \QUBO can be solved---exactly or approximately---with a wide range of optimization strategies.
A comprehensive overview can be found in \cite{kochenberger2014}.

The idea of embedding Sudoku in \QUBO in inspired by \cite{bauckhage2021a}, where an embedding is described for generalized Sudoku with side length $N^2$ and solved using Hopfield networks. 
The authors employ constraints to ensure that all numbers appear once in every row, column and block, that no cell is empty, and that the given clues are used correctly.
This work deviates from this strategy in two ways:
Instead of using additional constraints for clues, we fix a subset of variables and remove them from the optimization altogether, reducing the problem size considerably.
Further, we use no constraints to force all cells to be non-empty, but encourage non-empty solutions through reward and rely on the \QUBO solver to discover the correct solutions, which have lowest energy.

While \cite{bauckhage2021a} presents a general solution for any $N$, we focus on $N=3$ as the default Sudoku in this work.
However, a generalization of the techniques presented hereinafter is trivial.

\section{The Sudoku QUBO}

We introduce binary indicator variables $z_{ijk}\in\BB$ for all $i,j,k\in[9]$ with the interpretation ``the cell in row $i$, column $j$ has number $k$'', resulting in a total of $729$ variables that describe a Sudoku grid with every possible combination of numbers.
The constraints described in \cref{sec:intro} are encoded into a \QUBO weight matrix $\mat S$ through a positive penalty weight $\lambda$.
Given two cells $(i, j, k)$ and $(i', j', k')$, we have to penalize if any of four conditions hold: \begin{enumerate}[label=(\roman*)]
	\item $i=i' ~\wedge ~j\neq j' ~\wedge ~k=k'$\\(same number twice in the same row)
	\item $i\neq i' ~\wedge ~j=j' ~\wedge ~k=k'$\\(same number twice in the same column)
	\item $\floor{i/3}=\floor{i'/3} ~\wedge ~\floor{j/3}=\floor{j'/3} ~\wedge ~k=k'$\\(same number twice in the same block)
	\item $i=i' ~\wedge ~j=j' ~\wedge ~k\neq k'$\\(multiple numbers in the same cell)
\end{enumerate}

\begin{figure}
	\centering
	\sudoku{nyts}{0.8}
	\caption{Hard Sudoku puzzle from the New York Times on January 8, 2024, containing 24 clues. The solution is shown in \cref{sec:solution}.}
	\label{fig:nyts}
\end{figure}

Using only penalties would assign the same energy to empty cells without any bits set to 1 as to filled-out cells.
For this reason, we reward solutions for having a high number of 1-bits by giving each variable $z_{ijk}$ a linear weight of $-1$.
In order to represent all $z_{ijk}$ as a continuous bit vector, we need an arbitrary but fixed bijection $\iota: [9]^3\rightarrow[729]$, e.g., \begin{equation}
	\iota(i,j,k)\defeq 81i+9j+k\;.
\end{equation}
This allows us to work with bit vectors $\vec x\in\BB^{729}$ such that $x_u=z_{ijk}$ if $\iota(i,j,k)=u$ for all $u\in[729]$.
Finally, the entries of $\mat S$ can be summarized as \begin{equation}\label{eq:sudoqubo}
	S_{uv}\defeq\begin{cases}
		\lambda &\parbox[t]{5cm}{if $u\neq v$ and any of (i)-(iv) are true, where $u=\iota(i,j,k)$ and $v=\iota(i',j',k')$}\\
		-1 &\text{if }u=v\\
		0 &\text{otherwise.}
	\end{cases}
\end{equation}

The value of $\lambda$ has to be chosen such that it cancels out the reward of two placed numbers, which implies $\lambda>2$.
We simply set $\lambda=3$ for the remainder of this article.

\subsection{Incorporating Clues}
\label{sec:clues}

In a normal Sudoku puzzle, a subset of grid cells comes pre-filled with numbers that serve as clues, which the player uses to fill in the remaining numbers through deductive reasoning.
Let $C$ denote a set of tuples $(i,j,k)$ which are given as clues.
We can incorporate them by fixing the values of the corresponding variables at indices $\iota(i,j,k)$ and exclude them from the optimization procedure, reducing the search space size by half with every clue.

\paragraph{Clamping variables}

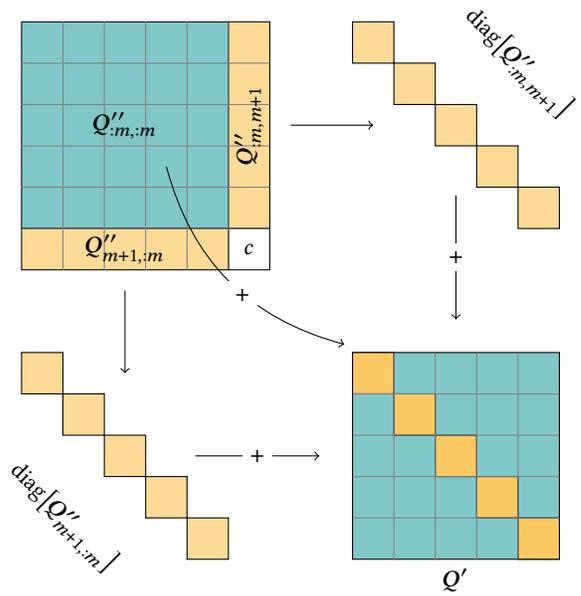
\begin{figure}
\begin{tikzpicture}[scale=0.55]
	\draw[fill=ml2rtrq!50!white] (0,1) rectangle (5,6);
	\draw[fill=ml2ryel!50!white] (0,0) rectangle (5,1);
	\draw[fill=ml2ryel!50!white] (5,1) rectangle (6,6);
	\draw (5,0) rectangle (6,1);
	\foreach \x in {1,2,3,4,5}{%
		\draw[color=gray] (\x,0) -- (\x,6);
		\draw[color=gray] (0,\x) -- (6,\x);
	}
	\node at (2.5,3.5) {$\mat Q''_{:m,:m}$};
	\node at (2.5,0.5) {$\vec Q''_{m+1,:m}$};
	\node[rotate=90] at (5.5,3.5) {$\vec Q''_{:m,m+1}$};
	\node at (5.5,0.5) {$c$};
	
	\draw[fill=ml2rtrq!50!white] (8,-7) rectangle ++(5,5);
	\foreach \x in {0,1,2,3,4}{%
		\draw[fill=ml2ryel!50!white] ($(8+\x,5-\x)$) rectangle ++(1,1);
		\draw[fill=ml2ryel!50!white] ($(\x,-3-\x)$) rectangle ++(1,1);
		\draw[fill=ml2ryel!75!white] ($(8+\x,-3-\x)$) rectangle ++(1,1);}
	\foreach \x in {0,1,2,3}{%
		\draw[color=gray] ($(8,\x-6)$) -- ++(5,0);
		\draw[color=gray] ($(\x+9,-7)$) -- ++(0,5);}
	\node[rotate=-45] at (12,5) {$\diag{\vec Q''_{:m,m+1}}$};
	\node[rotate=-45] at (1,-6) {$\diag{\vec Q''_{m+1,:m}}$};
	\draw[->] (6.5, 3.5) -- ++(2,0);
	\draw[->] (2.5,-0.5) -- ++(0,-2);
	\draw[->] (10.5,1.8) -- ++(0,-3) node[midway,fill=white] {$+$};
	\draw[->] (4.2,-4.5) -- ++(3,0) node[midway,fill=white] {$+$};
	\draw[->] (3.5,2.5) to [bend right] node[midway,fill=white,xshift=5pt,yshift=-5pt] {$+$} (7.8,-1.8) ;
	\node at (10.5,-7.5) {$\mat Q'$};
\end{tikzpicture}
\caption{Computation of $\mat Q'$ from $\mat Q''$.}
\label{fig:matrix}
\end{figure}

We can transform any \QUBO instance $\mat Q$ of size $n$ into a smaller instance $\mat Q'$ by assigning fixed values to one or more variables, sometimes referred to as \emph{clamping} \cite{booth2017}.
Assume we have two sets $I_0,I_1\subseteq [n]$ with $I_0\cap I_1=\emptyset$, and we want to implicitly assign $x_i=0 ~\forall i\in I_0$ and $x_j=1 ~\forall j\in I_1$.
Doing this allows us to eliminate these variables, such that $\mat Q'$ has only size $m=n-\abs{I_0}-\abs{I_1}$.
Firstly, we introduce a function $\kappa$ that maps the remaining variable indices $[n]\backslash I_0\backslash I_1$ to $[m]$.
Given a vector $\vec x'\in\BB^m$, we can compute the original energy value by re-inserting the fixed bits into $\vec x'$ and evaluating with $\mat Q$.
For simplicity, assume that $\vec x'$ is padded with an additional constant $1$ at the end, obtaining $\vec x''=(x'_1,\dots,x'_m,1)\in\BB^{m+1}$.
We can construct a matrix $\mat T\in\BB^{n\times(m+1)}$ that re-inserts the implicit bits, whose rows read \begin{equation}\label{eq:matT}
	\vec T_{i,\cdot}=\begin{cases}
		(0,0,\dots,0,0) &\text{if }i\in I_0,\\
		(0,0,\dots,0,1) &\text{if }i\in I_1,\\
		\vec{e}^\tp_{\kappa(i)} &\text{otherwise.}
	\end{cases}
\end{equation}%
The original energy value thus can be obtained as \begin{equation}\label{eq:clamped}
	f_{\mat Q}(\vec x) = \underbrace{(\mat T\vec x'')^\tp}_{\vec x^\tp}\mat Q\underbrace{(\mat T\vec x'')}_{\vec x}=\vec x''^\tp\underbrace{\mat T^\tp\mat Q\mat T}_{\mat Q''}\vec x''\;.
\end{equation}
The matrix $\mat T^\tp\mat Q\mat T\eqdef\mat Q''$ has size $(m+1)\times(m+1)$. However, we can reduce it to an $m\times m$ matrix by exploiting that $x''_{m+1}=1$, which lets us set $\mat Q'\defeq \mat Q''_{:m,:m}+\diag{\vec Q''_{m+1,:m}}+\diag{\vec Q''_{:m,m+1}}\in\mathbb{R}^{m\times m}$ due to $x_ix_{m+1}=x_i~\forall i\in [m]$.
Here, $:m$ as an index denotes all rows or columns up to and including index $m$, as illustrated in \cref{fig:matrix}.
The value $Q''_{m+1,m+1}\eqdef c$ is an additive constant, which can be ignored during optimization and carried along to recover the original energy value, which is \begin{equation}
	f_{\mat Q}(\vec x)=f_{\mat Q'}(\vec x')+c=\vec x'^\tp\mat Q'\vec x'+c\;.
\end{equation}

\par
Given our set of clues $C$, we can clamp the relevant variables of $\mat S$ the following way:
For all $(i,j,k)\in C$, \begin{enumerate}[label=(\Roman*)]
	\item\label{clamp1} clamp $x_{\iota(i,j,k)}=1$, fixing the correct value for cell $(i,j)$;
	\item\label{clamp2} $\forall k'\neq k$, clamp $x_{\iota(i,j,k')}=0$, removing the remaining possible values for cell $(i,j)$;
	\item\label{clamp3} $\forall i'\neq i, j'\neq j$, clamp $x_{\iota(i',j,k)}=0$ and $x_{\iota(i,j',k)}=0$, forbidding the same value in the same row or column;
	\item\label{clamp4} clamp $x_{\iota(i'',j'',k)}=0$ for all cells $(i'',j'')$ in the same block as $(i,j)$.
\end{enumerate}

The operations \labelcref{clamp2,clamp3,clamp4} are optional, but further reduce the size of the \QUBO instance.
Using operations \labelcref{clamp1,clamp2} alone, the number of variables is reduced by $9\abs{C}$.

A minimal Python implementation of \cref{eq:sudoqubo} and the clamping operations \labelcref{clamp1,clamp2,clamp3,clamp4} using the \verb+qubolite+ package\footnote{\url{https://smuecke.de/qubolite/index.html}} is given in \cref{sec:code}. 

\section{Practical Example}

We take the hard Sudoku puzzle from the New York Times website\footnote{\url{https://www.nytimes.com/puzzles/sudoku/hard}} from January 8, 2024, which is shown in \cref{fig:nyts}.
It has 24 clues, whose corresponding set $C$ is \begin{align*}
	C&=\lbrace(1,8,2),(2,5,9),(2,8,4),(3,4,3),(3,6,2),(3,9,7),\\
	&(4,1,6),(4,3,5),(4,4,1),(4,9,8),(5,4,7),(5,6,5),\\
	&(5,7,1),(6,2,3),(6,4,2),(6,9,5),(7,2,9),(7,5,7),\\
	&(7,9,3),(8,2,8),(8,8,9),(9,3,4),(9,6,6),(9,7,5)\rbrace\;.
\end{align*}%
This allows for a reduction to 513 variables using operations \labelcref{clamp1,clamp2}, and further down to 211 variables using \labelcref{clamp3,clamp4}.
The resulting matrix $\mat Q'$ is shown in \cref{fig:matrix}.
We eliminated two more variables using the QPRO+ preprocessing algorithm \cite{glover2018}, which is completely optional, though.
We solved the final \QUBO instance with 209 variables using Simulated Annealing (SA) and Quantum Annealing (QA).

\begin{figure}
	\centering
	\includegraphics[width=\columnwidth]{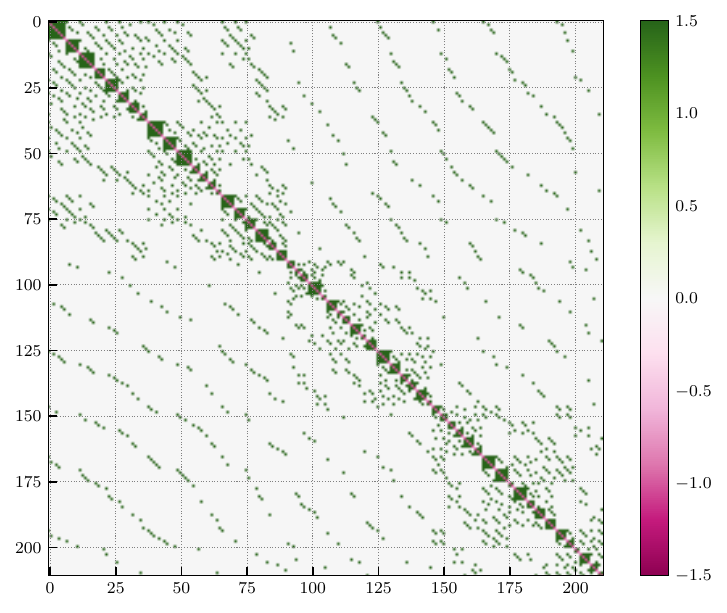}
	\caption{\QUBO parameter matrix for the hard Sudoku puzzle of the day by the New York Times on January 8, 2024.
		The image shows the symmetric matrix $0.5(\mat Q'+\mat Q'^\tp)$ after applying all clamping operations \labelcref{clamp1,clamp2,clamp3,clamp4}; the resulting optimization problem has 211 variables.}
\end{figure}

For SA, we used the \verb+SimulatedAnnealingSampler+ class from the \verb+dwave-neal+ package.
Similarly, for QA, we use the \verb+DWaveSampler+ class from \verb+dwave-system+.
For both, we set the number of readouts to 1000 and leave all other parameters untouched, so that they are set heuristically.

By construction, we know that \QUBO instance $\mat S$ from \cref{eq:sudoqubo} has a minimal energy of $-81$, which allows us to easily verify if a given solution $\vec x$ is correct:
If $\vec x^\tp\mat S\vec x=-81$, then $\vec x$ is a valid solution, as it contains all 81 numbers, and no constraints are violated.

SA is able to find the optimal solution quite reliably within 1000 readouts, with an average energy of $-75.047\pm 1.822$ per solution.
QA was \emph{not} able to find the correct solution, even after 15 tries with 1000 readouts each.
Its average solution energy is around $-53.795\pm 4.321$, which is quite far from the optimal $-81$.
The reason for this is probably \begin{enumerate*}[label=(\alph*)]\item the restricted topology of D-Wave's quantum annealing hardware, which leads to variable reduplication that increases the overall problem complexity, and \item a lack of manual hyperparameter tuning (e.g., chain strength, qubit mapping) \end{enumerate*}, which we were not feeling in the mood for.

As we achieve the best results with SA, we take a close look at its performance on a variety of Sudoku puzzles to gain insights into the problem's hardness depending on the number of clues given.

\subsection{Effect of Puzzle Difficulty on Performance}

\begin{figure*}
	\centering
	\includegraphics[width=\textwidth]{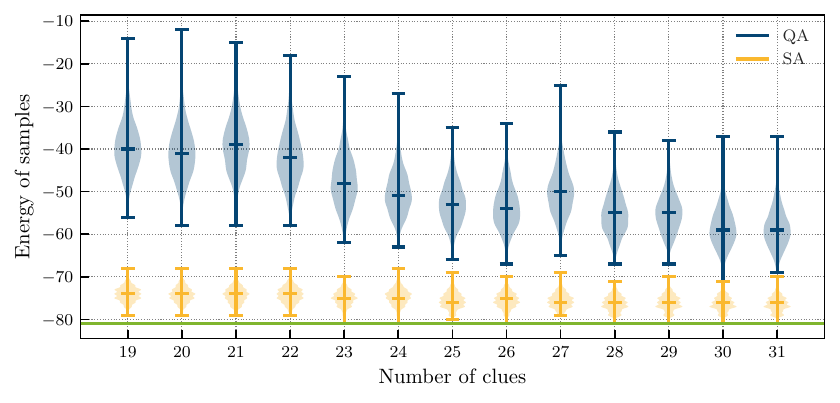}
	\caption{Sample energy distribution for 13 Sudoku puzzles of varying number of clues, using Simulated Annealing and Quantum Annealing with 2000 readouts each.
	The violin plots show minimum and maximum values as well as medians; lower energy is better.
	The green line is the lowest possible energy.}
	\label{fig:clues}
\end{figure*}

To this end, we use the ``3 million Sudoku puzzles with ratings'' data set created by David Radcliffe on Kaggle\footnote{\url{https://www.kaggle.com/datasets/radcliffe/3-million-sudoku-puzzles-with-ratings}}, which---faithful to its name---contains 3 million Sudoku games of varying difficulties.
Generally, Sudoku puzzles are harder the fewer clues they have, and the data set contains puzzles with numbers of clues $n_c$ ranging from 19 to 31.
For each of these values, we choose the first puzzle in the data set that has $n_c$ clues, giving us 13 puzzles which we again solve using both SA and QA.
This time, we perform 2000 readouts (the QA solver would not allow for higher values) and record the sample energies.
The results are visualized in \cref{fig:clues}.

Again, SA performs significantly better, finding the optimal solution for 7 out of 13 puzzles within 2000 samples, the hardest having 23 clues.
QA, on the other hand, is still not able to solve the problem with the given number of clues.
As we expected, puzzles with more clues are easier to solve for both methods, as the number of variables decreases with a higher number of clues.
A few more tries reveal that QA is able to solve puzzles with around 36 clues, when the \QUBO size approaches 100 variables.

\section{Sampling Sudokus}

In addition to solving Sudokus, the \QUBO formulation presented here can be used to sample uniformly from the space of valid Sudoku solutions.
As all of them have an energy value of $-81$, they are equally likely under the respective Gibbs distribution $p_{\mat S}(\vec x)=\exp\left[-\beta f_{\mat S}(\vec x)-A(\mat S,\beta)\right]$, where $\beta$ is the inverse system temperature and $A(\mat S,\beta)=\log\sum_{\vec x'\in\BB^n}\exp\left[-\beta f_{\mat S}(\vec x')\right]$ the log partition function.
For $\lim_{\beta\rightarrow\inf}$, the probabilities for all $\vec x'\in\BB^n$ with $f_{\mat S}(\vec x')>-81$ become 0, leaving---in theory---a uniform distribution over all optima.
However, on real quantum devices, tiny fluctuations of the parameter values, e.g., caused by integrated control errors, may bias the distribution towards a small subset of solutions (c.f. \cite{pochart2022}).

Using SA with a random initialization, on the other hand, circumvents these problems for now, allowing us to generate solved Sudokus very easily.
To test this, we took the full matrix $\mat S$ from \cref{eq:sudoqubo} and ran SA on it with $10,000$ readouts.
We found that $221$ of the obtained samples had an energy value of $-81$, meaning they are valid Sudoku solutions.
We further confirmed that all of them were distinct from each other.

Generating good Sudoku puzzles is more challenging, though:
While it is certainly possible to simply take a solved Sudoku, choose a subset of cells as clues and erase all other numbers, the resulting puzzle might have more than one solution, which is undesirable.
It was proven that the smallest number of clues necessary to obtain an unambiguous Sudoku puzzle is 17 \cite{mcguire2014}, therefore an algorithm to produce puzzles may look like this: \begin{enumerate*}[label=(\arabic*)]
	\item\label{item:1} Sample a solved Sudoku $s$,
	\item sample a set of clues $C$ with $\abs{C}\geq 17$,
	\item clamp $\mat S$ with $C$ to obtain $\mat S'_C$,
	\item draw $K$ samples from $\mat S'_C$; if any two samples $s'\neq s''$ both have energy $-81$, go back to \labelcref{item:1}, otherwise return $C$ as the puzzle.
\end{enumerate*}
The larger we choose $K$ the more confident we can be that $C$ has a unique solution.
However, it can still produce ambiguous puzzles, if by chance only one solution appears in the sample set.
Naturally, for $\abs{C}$ close to 17 this procedure may be unreliable, as the search space is vast, and the number of unambiguous Sudoku puzzles with such few clues is small.

\section{Summary \& Conclusion}

In this work we presented a simple way to encode the constraints of Sudoku into a \QUBO instance.
To solve a given puzzle, we showed how clues can be incorporated into the \QUBO weight matrix by clamping, which reduces the search space and aligns nicely with the intuition that Sudokus are harder if they have fewer clues.

We demonstrated that the method works as expected by performing experiments, both using a classical Simulated Annealing solver and a Quantum Annealer.
Our results show that SA can solve all presented test puzzles, while QA works only for rather easy puzzles at this point in time.
Nonetheless, it shows that new and emerging computing paradigms are ready for a wide range of computing tasks.
By reducing the problem size through implicit variables, we were able to solve a Sudoku with 35 clues on real quantum hardware for the first time, to the best of our knowledge.

Further, we showed that the \QUBO instance can be used for sampling valid Sudoku solution and sketched an algorithm for generating Sudoku puzzles with unique solutions, which is interesting for producing new puzzles.

\section*{Acknowledgments} 

This research has been funded by the Federal Ministry of Education and Research of Germany and the state of North-Rhine Westphalia as part of the Lamarr Institute for Machine Learning and Artificial Intelligence.

\appendix
\section{Solution to Puzzle}
\label{sec:solution}

This is the solution of the puzzle shown in \cref{fig:nyts}, found using the method described in this paper and Simulated Annealing.
\begin{figure}[H]
	\centering
	\sudoku{nyts-solved}{0.8}
\end{figure}

\newpage
\section{Python Code}
\label{sec:code}

The following Python function computes the Sudoku \QUBO weight matrix and returns it as a \verb+numpy+ array.

\begin{python}
import numpy as np
from qubolite import qubo # <- install via "pip install qubolite"
	
def compute_sudoku_qubo(lambda_=3.0):
    Q = np.zeros((729, 729))
    P = lambda_*(1-np.eye(9))-np.diag(np.ones(9))
    for u in range(81):
        Q[9*u:9*u+9, 9*u:9*u+9] = P
        i, j = u//9, u%9
        for v in range(u+1, 81):
            i_, j_ = v//9, v%9
            same_block = i//3==i_//3 and j//3==j_//3
            if i==i_ or j==j_ or same_block:
                for k in range(9):
                    Q[9*u+k, 9*v+k] = lambda_
    return qubo(np.triu(Q))
\end{python}

To apply the clamping operations described in \cref{sec:clues}, we use the \verb+partial_assignment+ class from \verb+qubolite+, which parses strings of the form \verb+x1=0; x2, x4=1+ to compute the matrix $\mat T$ (\cref{eq:matT}) and the reduced \QUBO instance $\mat Q''$ (\cref{eq:clamped}).

\begin{python}
from itertools import product
from qubolite.assignment import partial_assignment
	
def get_partial_assignment(clues):
	iota = lambda i, j, k: 81*i+9*j+k-91 # index map
	s = '' # partial assignment instructions
	for i, j, k in clues:
	    u = iota(i, j, k)
		# (  I) clamp correct value
		s+=f'x{iota(i,j,k)}=1;'
		# ( II) clamp incorrect values in same cell
		other_k = set(range(9)).difference((k,))
		s+=','.join([f'x{iota(i,j,k_)}' for k_ in other_k])+'=0;'
		# (III) clamp incorrect values in row and column
		other_i = set(range(9)).difference((i,))
		s+=','.join([f'x{iota(i_,j,k)}' for i_ in other_i])+'=0;'
		other_j = set(range(9)).difference((j,))
		s+=','.join([f'x{iota(i,j_,k)}' for j_ in other_j])+'=0;'
		# ( IV) clamp in same block
		block = [(i_,j_) for i_,j_ in product(
			range(3*(i//3),3*(i//3)+3),
			range(3*(j//3),3*(j//3)+3))]
		block.remove((i,j))
		s+=','.join([f'x{iota(i_,j_,k)}' for i_,j_ in block])+'=0;'
	return partial_assignment(s, n=729)
\end{python}

The resulting \verb+partial_assignment+ object can be applied to the Sudoku \QUBO matrix like this:
\begin{python}
	S = compute_sudoku_qubo()
	clues = [(1, 2, 3), ...]
	PA = get_partial_assignment(clues)
	S_, c = PA.apply(S)
	# S_: clamped QUBO instance,
	# c : constant offset
\end{python}
\end{document}